\documentstyle[epsf,epsfig,wrapfig]{elsart}
\begin{document}
\def\sc#1{$_{\mbox{\rule{0cm}{0.2cm}#1}}$}
\def\scm#1{_{\mbox{\rule{0cm}{0.2cm}#1}}}
{\vskip0pt\parskip=0pt\leftskip=20pt}
\begin{frontmatter}

\title{A New Method to Reconstruct the Energy
and Determine the Composition of Cosmic Rays
from the Measurement of Cherenkov Light
and Particle Densities 
in Extensive Air Showers}

\author{A.\,Lindner}

\address{University\,of\,Hamburg,
II.\,Inst.f.Exp.Physik,
Luruper\,Chaussee\,149, D-22761\,Hamburg, Germany,
E-Mail:\,lindner@mail.desy.de}

\begin{abstract}
A Monte-Carlo study to reconstruct energy and mass
of cosmic rays with energies above 300\,TeV
using ground based measurements of the
electromagnetic part of showers initiated in the atmosphere
is presented.
The shower properties determined
with two detector arrays measuring the air Cherenkov light and the
particle densities as realized at the HEGRA experiment
are processed to determine the energy
of the primary particle without the need of any hypothesis
concerning its mass.
The mass of the primary particle is reconstructed
coarsely from the same observables in parallel to the
energy determination.
\end{abstract}

\begin{keyword}

Cosmic rays, energy, composition, extensive air showers,
air Cherenkov light, HEGRA experiment

\PACS 96.40.De, 96.40.Pq, 13.85.Tp, 13.85.-t

\end{keyword}

\end{frontmatter}
\section{Introduction}
Despite experimental analyses of charged cosmic rays (CR) since more
than 80 years the origin and acceleration sites of CR remain 
uncertain.
A striking feature of the energy spectrum of CR is the so called 
``knee'' around 3\,PeV.
Below the ``knee'' the differential energy spectrum follows
a simple power law with an index of $\approx$\,-2.7 over an energy range of
four orders of magnitude, above the knee the index decreases to
$\approx$\,-3.0 and stays
roughly constant for the next three orders of magnitude.
It is hoped that detailed measurements of the energy spectrum and 
the composition of CR around the knee may provide a clue to understand 
the origin of charged cosmic rays.
Possibly the ``knee'' indicates a transition between two different
classes of cosmic accelerators.  
Contemporary theoretical models describe the acceleration
of nuclei in the cosmos mostly by strong shocks \cite{shocks},
either of galactic or extragalactic origin,
which can be effectively produced in supernova remnants,
supersonic stellar winds, pulsar driven nebulae,
active galactic nuclei, relativistic jets in radio galaxies
and other phenomena \cite{theo}.\\
But neither the acceleration sites could be identified directly
by looking for photons resulting from interactions of the 
accelerated nuclei with the ambient medium, nor were measurements 
of the charged cosmic rays conclusive.
The latter results suffer seriously from the fact, 
that due to the low flux of CR above 1\,PeV
only large ground based installations observing the extensive air showers
(EAS) induced by cosmic rays in the atmosphere
provide experimental data in the energy regime around the ``knee''.
The sensitivity of EAS observables to the mass of
the primary particle is in general rather weak due primarily
to large fluctuations of the showers' developments in the atmosphere
which may be larger than mean differences between
primary cosmic proton and iron nuclei.
The interpretation of experimental data is rendered even more difficult
due to theoretical uncertainties concerning high energy interactions
in the atmosphere \cite{fzka5828,thprobs}.
Consequently the experimental results are not conclusive at
present and in some cases contradicting to each other
(see for example the reviews \cite{exrev}).
Therefore it is highly desired to develop new techniques 
to measure the properties of cosmic rays
between 1 and 10\,PeV.
Ideally one would combine detailed measurements of the hadronic 
and muonic shower properties (as performed by 
the experiment KASCADE \cite{kascade})
and precise analyses of the development of the electromagnetic 
shower component, which can be deduced from registered air 
Cherenkov light for example.
\\
This paper describes a new method to derive the energy spectrum
and the elemental composition
for energies above a few hundred TeV
from measurements of Cherenkov light and particle
density at ground level using an experimental setup as realized at
the HEGRA installation \cite{hegra,perpig}.
It will be shown that the energy of a primary particle can be
determined independently of its mass with an accuracy of around 10\%
in the knee region.
The mass of the primary particle can be reconstructed coarsely,
allowing the determination of the relative amount of light, medium
and heavy nuclei in cosmic rays without requiring any hypotheses
concerning their energy spectrum.
These results will be achieved by analyzing the electromagnetic part
of the air shower only.
Therefore the new method is complementary to many other results
concerning the experimental measurements and the theoretical
interpretations of data.
Preliminary results of an application of this method to HEGRA
data have been presented at conferences
(\cite{perpig,romax}
 and a similar method \cite{romra})
proving the validity of this composition analysis
by comparison with the results of direct experiments below 1\,PeV.\\
In contrast to earlier analyses of the air Cherenkov light produced in EAS
and results on the development of air showers
(see \cite{oldcheren} and references therein),
this MC study focuses on an interpretation of combined particle and
Cherenkov light measurements.
A detailed analysis of the Cherenkov light registered relatively close
to the shower core is the basis for the energy and mass reconstruction,
which can only be improved significantly by adding observables related
to the non electromagnetic shower components.\\
The paper is organized as follows:
Next the experimental observables which will be used to
reconstruct energy and mass of the primary particle are described.
Most important for this analysis are the shape of the lateral 
Cherenkov light density distribution, 
the amount of registered Cherenkov light and the number of charged particles
at detector level.  
In the third chapter 
the event simulation is sketched and features of the longitudinal and
lateral shower development in the atmosphere are considered in section four.
The following two sections deal with the reconstruction of
the position of the shower maximum from the shape of the Cherenkov 
light density distribution and with an estimation
of the energy per nucleon from the penetration depth of 
the shower until the maximum is reached.
In the 7${\rm ^{th}}$ section the primary energy is derived.
Here the shower size at detector level and the known position of 
the shower maximum are used to determine the shower size at the maximum.
This is combined with the energy per nucleon 
to calculate the primary energy 
independent of the primary mass.
The 8${\rm ^{th}}$ section presents methods to determine the chemical
composition from the observables related to the electromagnetic 
shower component.
Section\,9 discusses systematic uncertainties related to the simulation
code and estimates the sensitivity of the method presented in this paper
to different interaction models.
This is followed by a summary and conclusions.
\section{The experimental Observables}
Although the method presented in this paper was developed primarily for
the HEGRA experiment it can equally well be applied to any installation
registering air Cherenkov light and charged particles of extensive air
showers (EAS). The method can be easily generalized to all experimental
setups which allow the determination of the distance to the shower
maximum. However some properties of the HEGRA experiment need to be
mentioned to understand details discussed in the following sections.\\
The experiment HEGRA is a multi-component detector complex
for the measurement of EAS described elsewhere \cite{hegra}.
At a height of 2200\,m a.s.l.\ it covers an area of
180${\rm \cdot 180\,m^2}$.
In this paper only the scintillator array of 245 huts
on a 15\,m grid spacing and the so called
AIROBICC array of 77 open photomultipliers sampling the Cherenkov light
front of air showers on a 30\,m grid spacing are used.
Both components feature an increased detector concentration around
their common center.
The energy threshold (demanding a signal from
at least 14 scintillator or 6 AIROBICC huts) lies at 20\,TeV for proton
and 80\,TeV for iron induced showers.\\
\label{observables}
The measured particle density in the plane perpendicular to the
shower axis is fitted by the NKG formula \cite{nkg}.
In the fit a fixed Moliere radius of 112\,m is used.
The shape parameter {\em age} and the integral number of
particles {\em $N_e$} result from the fitting procedure.
As the HEGRA scintillators are covered with 5\,mm of lead (which suppresses
the detection of low energy electrons but
allows the measurement of photons after pair production in the lead)
the values obtained for {\em age} and {\em $N_e$} cannot be compared to
simple expectations
from the cascade theory.\\
The Cherenkov light density is analyzed in the interval
20\,m ${\rm <}$ r ${\rm <}$ 100\,m from the shower core,
because the shape of this part of the lateral distribution
appears to be independent of the mass of the primary
particle (see section\,\ref{recmax}).
In the range between 20 and 100\,m the Cherenkov light density
can be well described by an
exponential
\begin{equation}
\label{slopedef}
{\rm \rho_C(r)={\em a} \cdot exp(r\cdot {\em slope})}.
\end{equation}
Similar to the NKG fit two parameters are obtained from the analysis
of the Cherenkov light:
the shape parameter {\em slope}
and the total number of Cherenkov photons
reaching the detector level between 20 and 100\,m core
distance, {\em L(20-100)}.
\section{The Generation of the Event Sample}
EAS in the energy range from 300\,TeV to 10\,PeV were simulated using
the CORSIKA 4.01 code \cite{cors}.
The model parameters of CORSIKA were used with their default values
and the fragmentation parameter was set to ``complete fragmentation''.
This results in a complete disintegration of the nucleus after
the first interaction.
Showers induced by the primary proton,
${\rm \alpha}$, oxygen and iron nuclei were calculated.
The number of generated Cherenkov photons
corresponds to the wavelength
interval between 340 and 550\,nm.
In total 1168 events were generated with CORSIKA\,4.01
with zenith angles of 0,15,25 and 35$^0$\ at
discrete energies between 300\,TeV and 10\,PeV. 
Systematic comparisons of different event generators
to model the interaction of cosmic rays with nuclei of the
atmosphere will be sketched in section\,\ref{wo}.\\
In the main this paper assumes perfect measurements
of the number of particles and Cherenkov photons
and a perfect shower core determination
in order to concentrate on the physical principles and
limitations of the methods to be described.
To study the influence of the realistic experimental performance
the events were passed through a
carefully checked detector simulation \cite{volker}
and reconstructed
with the same program as applied to the real data.
Here each event was used 20 times to simulate
different core positions inside and impact points outside
the experimental area,
which nevertheless fire a sufficient number of detector huts to
fulfill the trigger conditions.
\section{The Development of Showers in the Atmosphere}
Some basic characteristics of the EAS simulated with
CORSIKA\,4.01 are summarized here.
Features independent as well as sensitive to the mass of the
nucleus hitting the atmosphere are described.
These will allow the reconstruction of
the primary energy and mass from the observables mentioned above.
\subsection{The Shape of the longitudinal Shower Development}
\label{long}
\begin{figure}[htb]
     \begin{center}
     \epsfxsize=13cm
     \epsfysize=6.16cm
     \mbox {\epsffile{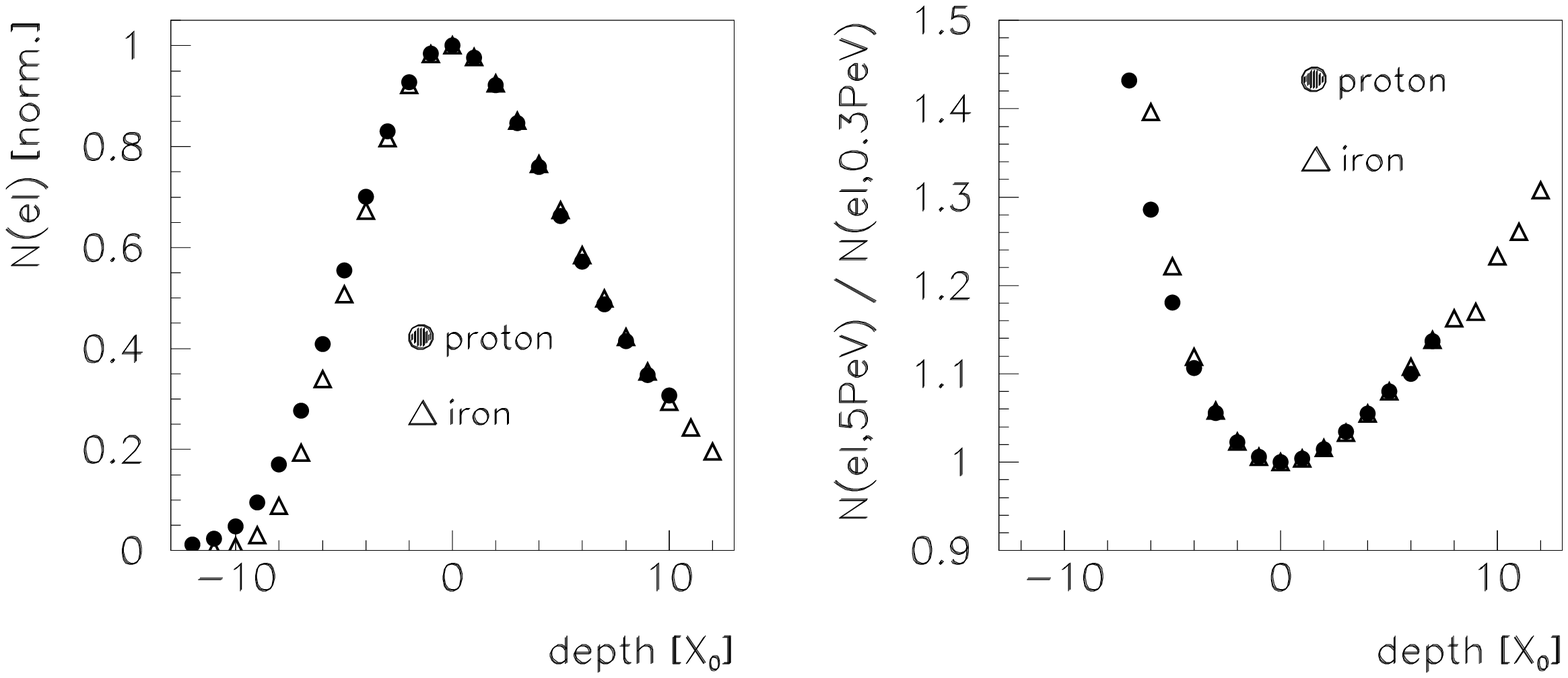}}
     \caption{Left: the mean longitudinal
development of 300\,TeV p and Fe
showers normalized to the number of particles at the maximum.
Not that the depth of each individual shower maximum has been
shifted to 0\,${\rm X_0}$ before averaging.
Right: the ratio of 
the mean longitudinal development for
5\,PeV showers divided by the mean
development for 300\,TeV showers (normalized at the
shower maximum at 0\,X$_0$).}
     \label{longdev}
     \end{center}
   \end{figure}

As discussed later in section\,\ref{recmax} 
the shape of the longitudinal shower development behind the
shower maximum is most important for 
the determination of the position of the shower maximum from 
the lateral Cherenkov light density distribution.
Shown on the left of Figure\,\ref{longdev} are
the mean longitudinal developments
of 300\,TeV proton and iron induced air showers, where
electrons and positrons above an energy of 3\,MeV were
counted.
This will be subsequently called, the shape of
the longitudinal shower development.
Note that for each shower the maximum (defined as the point in the shower
development with the maximal number of particles)
was shifted to zero before
averaging.
Afterwards the mean distribution was normalized to the mean particle number
at the shower maximum.
With regard to the shape of the longitudinal
development behind the shower maximum,
no systematic differences depending on the primary particle are visible.
The right plot in Figure\,\ref{longdev} shows the change in the
longitudinal development with increasing primary energy.
Independent of the primary particle the longitudinal
shower shape becomes more extended behind the
shower maximum with rising energy.
The independence of the longitudinal shower shape on
the mass of the primary nucleus
may be explained by a fortunate combination of two effects:
\begin{enumerate}
\item As visible from simulated proton
and iron showers at 300\,TeV and 5\,PeV
the longitudinal shower shape extends with increasing
energy (Figure\,\ref{longdev}, right).
\item After the first interaction an iron induced shower
can be described as a superposition of
nucleon induced subshowers.
Each of them has a different subshower maximum position,
which fluctuates around a mean value.
The longitudinal shape of the whole air shower
results from the overlay of all subshowers.
Therefore a Fe shower
appears to be longitudinally more extended than
a proton shower of the same energy per nucleon.
\end{enumerate}
Both effects combine in such a way that the longitudinal
shape of the EAS behind the maximum
becomes independent of the mass of the primary nucleon for a fixed
primary energy in the simulations used here.
This independence is 
most important for the methods discussed in the following sections. 
It will allow to determine the distance between detector and shower 
maximum with the same algorithm for all nuclei. 
\subsection{The Depth of the Shower Maximum}
\begin{figure}[htb]
     \begin{center}
     \epsfxsize=13cm
     \epsfysize=6.16cm
     \mbox {\epsffile{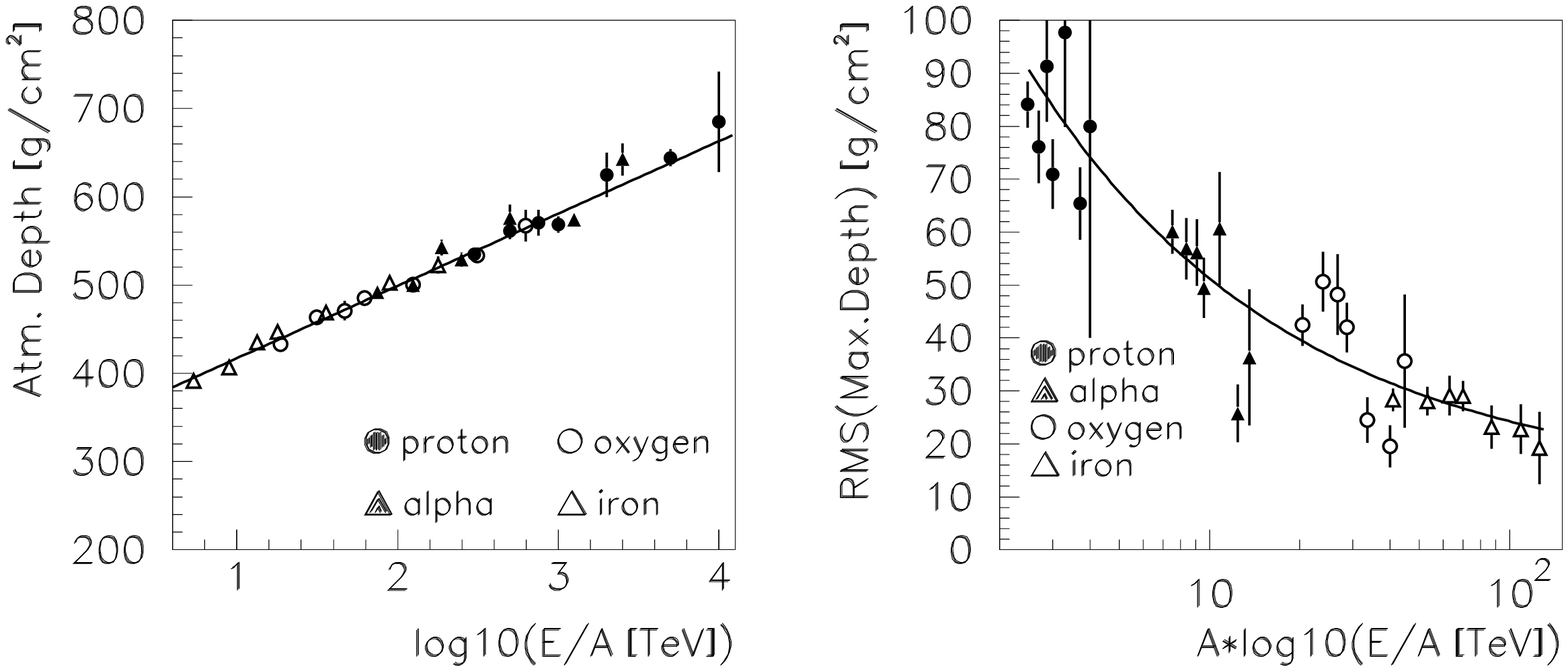}}
     \caption{Left: the mean atmospheric depth of the
shower maxima as a function of energy per nucleon.
The line shows a fit to the correlation.
Right: The fluctuations (rms) of the  atmospheric depth of the
shower maxima are plotted as a function of
energy E and nucleon number A.
The line shows a fit to the correlation.}
     \label{detenuc}
     \end{center}
   \end{figure}

The mean atmospheric depths of the maxima depend on the energy per nucleon
E/A and are subjected to large fluctuations.
Figure\,\ref{detenuc} (left) shows the corresponding correlation:
the column density traversed by a shower up to its maximum,
named depth of maximum in the following
(calculable from the distance and the zenith angle), is correlated
equally with E/A for all different simulated primaries from p to Fe
and all zenith angles.
With the ``complete fragmentation'' option in our simulations
the correlation follows a linear function.
From Figure\,\ref{detenuc} (left) an elongation rate of approximately
82\,g/cm$^2$/log$_{10}$(E/E$_0$) is derived:
\begin{equation}
{\rm depth(max)\,=\,\left[ (335 \pm 3) \,+\,(82 \pm 2) \cdot log_{10}
\left(\frac{E/A}{TeV}\right)\right]\,g/cm^2 .}
\label{eqenuc}
\end{equation}
If the depth of the shower maximum is measured the
energy per nucleon E/A can be inferred, but
due to statistical fluctuations of the depth of the
shower maxima (Figure\,\ref{detenuc}, right)
the resolution for E/A is modest.
The fluctuations decrease with increasing nucleon number A
as the EAS of a complex nucleus consists of
many overlapping nucleon induced subshowers so that the
whole EAS exhibits less variations
than the individual subshowers.
The fluctuations diminish slightly also with
rising E/A because more interactions take place
until the shower maximum is reached.
In addition it is interesting to note that for a specific 
primary particle and energy the number of particles in
the shower maximum Ne(max) does not depend 
significantly on the depth of the maximum.\\
The important features used in the following sections are:
\begin{itemize}
\item The mean depth of the shower maximum is determined only by E/A.
\item Fluctuations in the position of the shower maximum
decrease with increasing nucleon number.
\end{itemize}
\subsection{The lateral Shower Development}
\label{showerlat}
In hadronic interactions the typical transverse momentum
increases only very slowly with rising momentum transfer.
Therefore the lateral spread of
hadronic showers should decrease with increasing energy
per nucleon as
the ratio of transverse to longitudinal momentum gets smaller
in the early part of the shower development
where the energies of the interacting particles are still
comparable to the primary energy.
In principle this effect could be used as a measure for E/A
(if the distance to the shower maximum is known)
i.e.\ by comparing
the number of Cherenkov photons reaching the detector
level relatively close to the core with all photons
detectable at the ground level.
    \begin{figure}[bth]
     \begin{center}
     \epsfxsize=13cm
     \epsfysize=6.16cm
     \mbox {\epsffile{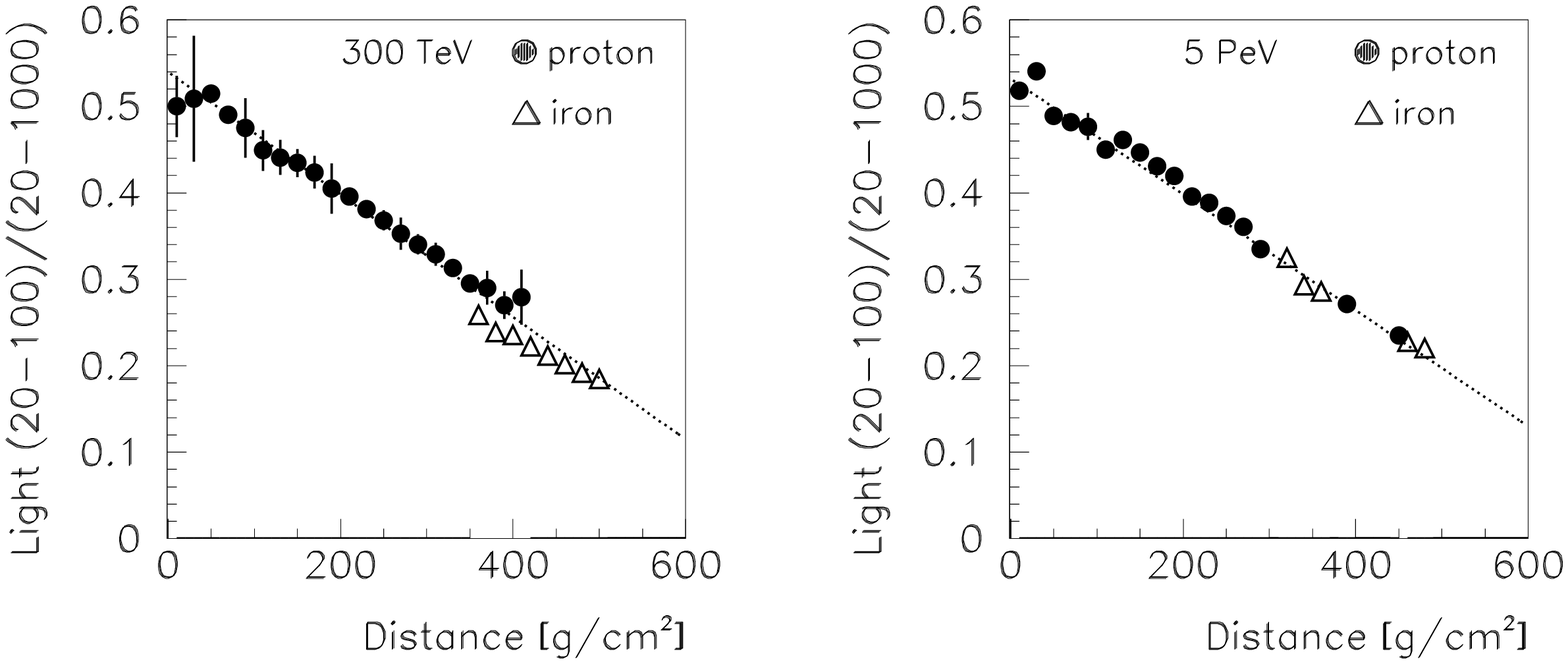}}
     \caption{The mean ratio of the Cherenkov light
reaching the detector level between 20 and 100\,m core
distance to all photons up to 1000\,m
for 0.3\,PeV (left) and 5\,PeV (right) iron and proton showers
as a function of the distance between detector and shower 
maximum.
The error bars show the rms values of the ratio at a fixed distance,
the lines are fits to the correlations for protons
to guide the eye.}
     \label{complight}
     \end{center}
   \end{figure}
In Figure\,\ref{complight} a distinction between heavy and light 
primaries is obvious at energies below 1\,PeV:
iron induced showers are broader than showers of primary protons.
However at energies in the knee region nearly no differences
between proton and iron showers remain.
Obviously here E/A, which determines the longitudinal
momentum in hadronic interactions,
gets so large even for iron showers that
any influence of the hadronic transverse momentum
in the first interactions becomes negligible
and the lateral extension of the shower is dominated by
scattering processes and interactions of
particles of lower energies
in the later part of the shower development.
Qualitatively the same result is achieved 
by correlating {\em age} with the distance 
to the shower maximum. 
Therefore these features of the lateral shower development 
can be used only for energies below 1\,PeV to enhance the 
mass sensitivity of EAS measurements in order to improve
comparisons with direct balloon data. 
\section{Reconstruction of the Distance to the 
        Shower Maximum}
\label{recmax}
This section deals with the reconstruction of the distance between
the shower maximum and the detector.
It will be shown that the distance to the shower maximum can
be determined from
the shape of the lateral Cherenkov light density with an
accuracy of about one radiation length (X$_0$) at 300\,TeV
and better than 0.5\,X$_0$ above 1\,PeV. \medskip \\
\label{recmaxcl}
\subsection{The Principle}
As already noticed by Patterson and Hillas \cite{hillas}
for proton induced showers with energies above 1\,PeV
the distance between the detector and the maximum of an EAS
can be inferred from the lateral distribution of the Cherenkov light
within about 100\,m core distance.
Simulations with CORSIKA show that this is possible independent of
the mass of the primary particle. \\
The basic ideas of using the lateral extension of
Cherenkov light to determine the position of the
shower maximum are the following:
Cherenkov light emitted at a specific height
during the shower development in the atmosphere shows
a specific lateral distribution at detector level.
The light from the early part of the shower development,
where the energies of the particles are still very high
so that scattering angles are very small,
is concentrated near 120\,m
(the so called Cherenkov ring at an observation level of 2200\,m).
Cherenkov light produced closer to the detector level
hits the ground closer to the shower core.
The measurable Cherenkov light density of the whole EAS
is the sum of all
contributions from all depth, where lateral distributions from different
depths enter with amplitudes corresponding to the number of
Cherenkov light emitting particles in the different depths.
Hence the shape of the measurable lateral light density
distribution depends
on the longitudinal shower development.
If the shower maximum approaches the observation level
more light is produced close to the detector reaching the
ground near the shower core.
Consequently the lateral Cherenkov light density
in the range up to 100\,m core distance
drops more rapidly,
if the shower maximum approaches the detector more closely. 
\vspace*{-0.5cm}\\
\begin{wrapfigure}[19]{r}{6.5cm}
\epsfig{file=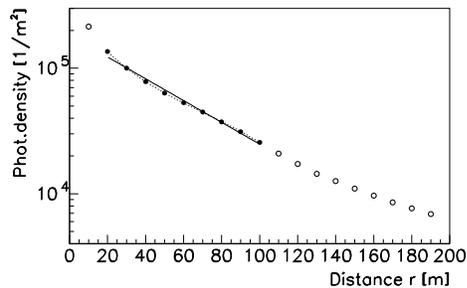,width=7cm,height=4.7cm,angle=0}
\caption{The lateral
Cherenkov light density of one simulated
300\,TeV proton shower
without experimental uncertainties as a function of the 
distance to the shower core.
Also shown are a fitted
exponential (full line, eq.\,\ref{slopedef}) and a function with
four parameters (dotted).}
\label{clattheo}
\end{wrapfigure}
It is important to note that the Cherenkov light in the discussed 
core distance range is radiated dominantly by particles behind 
the shower maximum. Therefore the longitudinal shower development 
behind the maximum determines
the correlation between the distance to the 
shower maximum and the shape of the lateral Cherenkov light density.
Figure\,\ref{clattheo} shows
a detailed view of the lateral Cherenkov light
distribution for one 300\,TeV proton shower.
A simple exponential fit can be used
to parameterize the distribution between 20 and 100\,m
core distance (equation\,\ref{slopedef}),
although this ansatz
cannot account for all details of
this distribution.
But taking into account a realistic experimental spread
of the light density measurements
it is unreasonable to determine more than {\em slope}
and an amplitude parameter in a fit.
Therefore just the simple exponential function is used
throughout this paper\footnote{If the 
distance to the maximum is determined using
the 4-parameter fit 
the resolution for the position of
the shower maximum compared to the method described below
improves by 30\% at 300\,TeV, whereas no difference in the
achievable accuracy was noted above 1\,PeV (if 
no experimental uncertainties are considered).}.
\subsection{The Distance to the Maximum}
\begin{figure}[htb]
     \begin{center}
     \epsfxsize=13cm
     \epsfysize=6.16cm
     \mbox {\epsffile{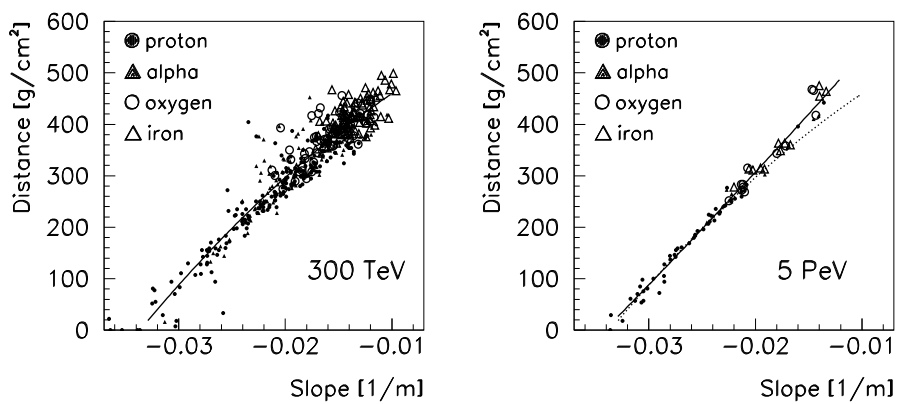}}
     \caption{ The distance between detector and
shower maxima plotted against the parameter {\em slope}
for 300\,TeV (left) and 5\,PeV (right) primary energy.
The lines show fits to the correlations.
The dotted line in the right plot corresponds to the
fit to 300\,TeV showers to visualize the
differences between different primary energies.
No experimental uncertainties are considered.}
     \label{maxcor}
     \end{center}
   \end{figure}

In Figure\,\ref{maxcor} the correlation between the distance to
the shower maximum and the parameter {\em slope} derived from the
Cherenkov light distribution is plotted for different primary
nuclei and zenith angles up to 35$^\circ$
for primary energies of 0.3 and 5\,PeV.
The distance to the shower maximum can be determined
from {\em slope}
independent of the type of the primary particle and
zenith angle\footnote{neglecting atmospheric absorption}.
The correlation between the distance to the shower maximum
and {\em slope} depends slightly on the
primary energy due to the changing
longitudinal shower development (see section\,\ref{long}).
Figure\,\ref{maxcor} shows simple polynomial fits describing
the correlation rather well.
The dependence of the fit parameters on log(E) were again
parameterized with polynomials
resulting in a two dimensional function of
{\em slope} and log(E) to determine the distance to the shower
maximum.
The systematic uncertainties in the reconstruction of the
distances
for primary nuclei from p to Fe and for different
primary energies are less than 5\%
increasing a little for zenith angles of 35$^\circ$.\\
The accuracy of the
determination of the distance to the shower maximum
(for a given primary energy)
improves with decreasing distance between
detector and shower maximum and with increasing
number of nucleons.
      \begin{figure}[tbh]
     \begin{center}
     \epsfxsize=13cm
     \epsfysize=6.16cm
     \mbox {\epsffile{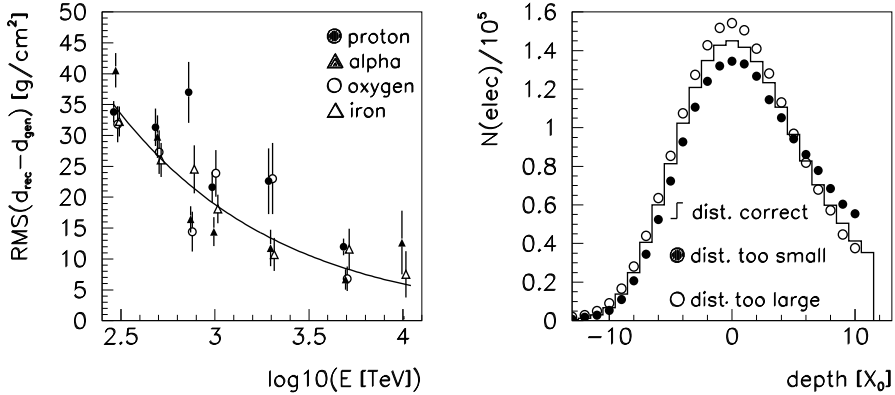}}
     \caption{Left: the rms value of the distributions
of the absolute difference between reconstructed
(d{\footnotesize rec})
and MC generated distance (d{\footnotesize gen})
to the shower maximum for
zenith angles of 0$^0$ and 15$^0$ as a function of energy E.
The line shows a fit to the correlation.
The rms values increase for larger zenith angles
(not shown).
Right: the mean longitudinal development of 300\,TeV proton
showers where the reconstructed distance is more than
20\,${\rm g/cm^2}$ too large (open dots), too small (full dots)
or correct within 10\,${\rm g/cm^2}$ (line).
The maximum for each individual shower was shifted to
zero before averaging.
In contrast to Figure\,\ref{longdev} the distributions were not
normalized to the number of particles in the maximum.}
     \label{distrms}
     \end{center}
   \end{figure}
Both contributions compensate such that
the resolution averaged over all distances becomes
independent of the mass of the primary particle within the
statistical errors of the event sample.
The resulting precision of the distance determination
is plotted in Figure\,\ref{distrms} (left) as a function
of the primary energy.
Fluctuations of the shape of the longitudinal shower development limit
the resolution of the distance determination with {\em slope},
which is shown in the right part of Figure\,\ref{distrms}. 
Showers, where the distance is underestimated (overestimated), 
do not decay as fast (faster) as an average shower behind the 
shower maximum.
Therefore {\em slope} is smaller (larger) than for the mean
longitudinal shower development and the 
distance is reconstructed too small (large).
Figure\,\ref{distrms} (right) furthermore proves, that,
as expected, the length of the shower behind the
maximum is anti correlated with the number of particles in the
maximum: the faster the decay after the maximum the more particles
arise in the maximum. \\
Further MC studies show that a method to determine the distance
to the shower maximum from the lateral Cherenkov light distribution
independent of the mass of the primary nucleus
is only possible if the analysis is restricted to core distances
below the so-called Cherenkov ring at 120\,m (at an observation
level of 2200\,m).
Extending the analyzed region beyond 120\,m introduces
a particle dependent bias (compare Figure\,\ref{complight}),
which however becomes negligible for energies in the region of the knee.
Of course other approaches,
like the reconstruction of the shower maximum from measurements of the
time profile of the Cherenkov light pulses
at core distance beyond 150\,m
\cite{volker} show different limitations as
different shower properties are measured.
\medskip \\
In this section it was shown, that the distance between detector and
shower maximum can be derived from the shape of the lateral
Cherenkov light density independent of the primary mass.
Taking into account a weak dependence on the primary energy
a resolution of better than 0.5 radiation lengths is achieved
for energies in the knee region.
\section{Determination of the Energy per Nucleon E/A}
\label{detea}
\begin{wrapfigure}[24]{r}{6.5cm}
\epsfig{file=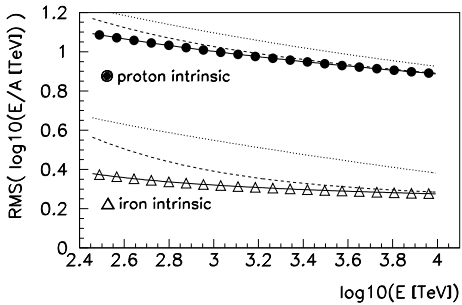,width=7cm,height=4.7cm,angle=0}
\caption{The accuracy of the
E/A reconstruction:
``intrinsic'' contributions refer to Figure\,\ref{detenuc}
showing the natural fluctuations of the shower maxima positions.
The broken lines displays the accuracy
achievable with the method
to reconstruct the distance to the shower maximum with {\em slope}
assuming perfect measurements of the Cherenkov light densities,
the dotted lines shows the results, when a
realistic performance of the present HEGRA
detector is included
(0 and 15$^0$\ zenith angle).}
\label{enucrms}
\end{wrapfigure}
With known distance to the shower maximum and known zenith angle
the penetration depth of the shower into the atmosphere until it
reaches the maximum can be inferred.
The energy per nucleon E/A can be estimated
using the correlation shown in Figure\,\ref{detenuc} (left).
In Figure\,\ref{enucrms} the accuracies
of the E/A measurement are shown assuming a
perfect determination of the depth of the
shower maximum (Fig.\,\ref{detenuc})  
as well as deriving the depth from an
ideal (Fig.\,\ref{distrms}) 
and realistic measurement of {\em slope}.
The uncertainty of the E/A determination for
proton induced showers is always dominated by statistical
fluctuations of the shower maximum position,
whereas for iron showers at low energies the intrinsic uncertainty
of the {\em slope} method contributes significantly.
At energies above the ``knee'' the accuracy is limited
by variations of the shower maxima positions for all primary nuclei.
\section{Reconstruction of the primary Energy}
The algorithms to reconstruct the primary energy of cosmic rays
described in the present paper can be roughly divided into
two steps:
first the distance of the shower maximum to the detector (derived
from the shape of the lateral Cherenkov light density distribution)
is combined with the measured number of particles or
Cherenkov photons at detector level.
In such a way the dependence of these two quantities on different
shower maximum distances can be corrected for and an
accurate energy determination becomes possible in spite of
large natural fluctuation in the position of the shower maxima.
Especially the shower size at the maximum Ne(max) will be 
determined from the shower size at detector level and {\em slope}. \\ 
As only experimental quantities measuring
the electromagnetic part of the air shower are considered here
it follows naturally, that only the energy
deposited in the electromagnetic cascade can be reconstructed
directly.
In a second step a correction for the non measured energy
has to be performed.
This correction depends on E/A only, which is determined from the
depth of the shower maximum as described in the previous section.\\
The following plots and parameterizations
only take into account showers which reach their maxima at least
50\,g/cm${\rm^2}$ above the detector, otherwise
one can hardly decide whether a shower reaches its maximum
above detector level at all.
The treatment of showers arriving at detector level before reaching
their maxima has to be considered separately.
\subsection{Energy Reconstruction from Particle and
Cherenkov Light Measurement}
\label{ene}
\begin{wrapfigure}[16]{r}{6.5cm}
\epsfig{file=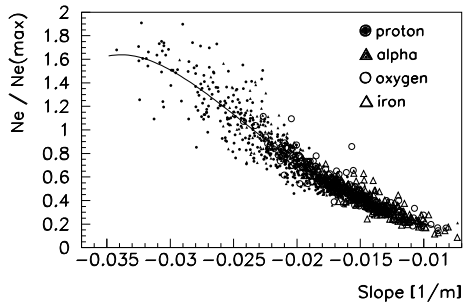,width=7cm,height=4.7cm,angle=0}
\caption{The ratio of {\em $N_e$} (measured below 0.5\,cm of
lead) and the
number of particles at the maximum Ne(max)
as a function of {\em slope}.
The line shows a fit to the correlation,
named $\xi {\rm _{dis}}$({\em slope}).}
\label{eemcor}
\end{wrapfigure}
\begin{figure}[htb]
     \begin{center}
     \epsfxsize=13cm
     \epsfysize=6.16cm
     \mbox {\epsffile{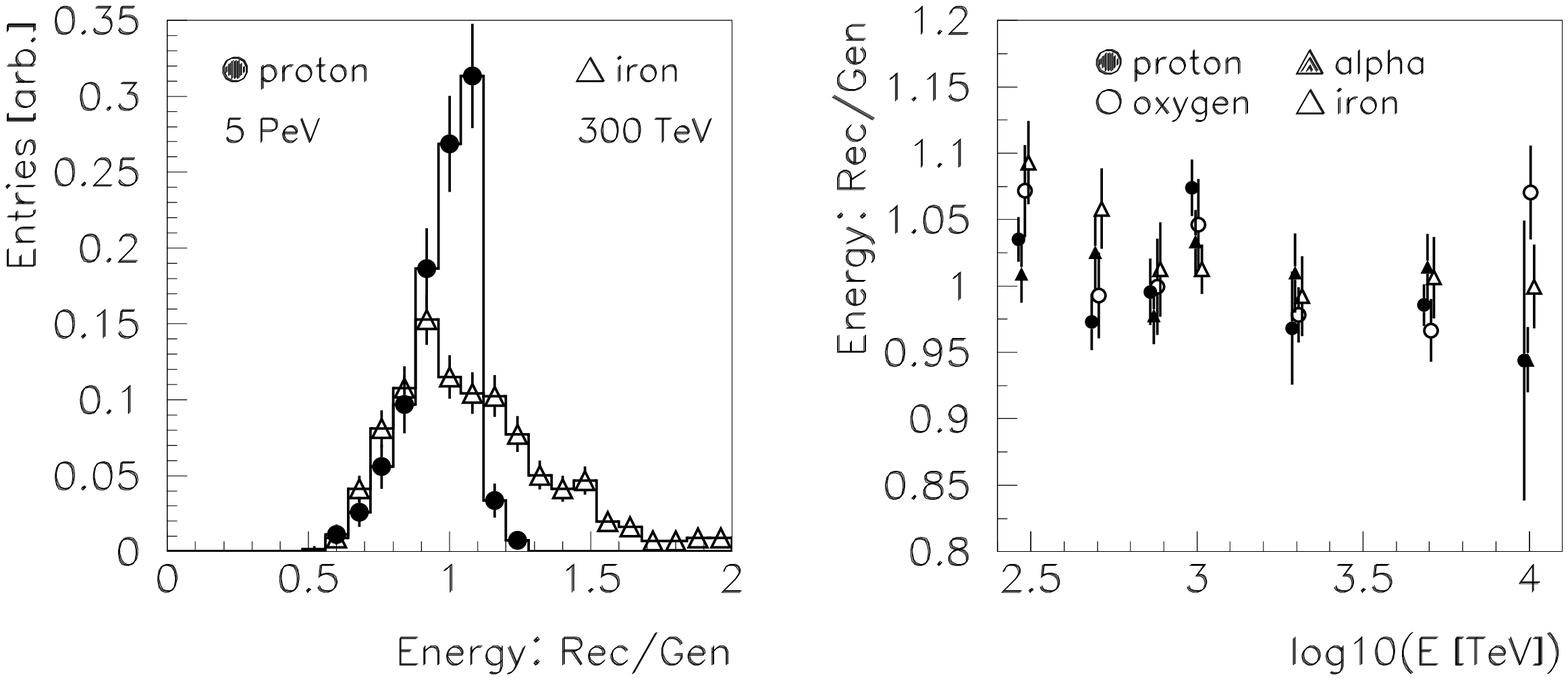}}
     \caption{Left: the ratio of reconstructed
and generated energy for 5\,PeV proton
and 300\,TeV iron showers (as two extreme cases for 
the energy per nucleon in the used event sample).
Right: the mean value of the same ratio for
different primaries and energies (right).
Only events with a minimal distance of ${\rm 50\,g/cm^2}$
were considered in both plots.
Each generated event was used several times with different
core positions to take into account the experimental uncertainties
for the {\em $N_e$} determination properly.}
     \label{etotaccu}
     \end{center}
\vspace*{0.5cm}
   \end{figure}

The first step for reconstructing the primary energy
is the determination of the number of particles in the 
shower maximum Ne(max) from the observables
{\em $N_e$} and {\em slope},
because Ne(max) is a good measure of the energy contained 
in the e.m.\ cascades.\\
As the scintillator huts of the HEGRA experiment are
covered with 1\,${\rm X_0}$ of lead the total number of
particles is not measured directly. In simulations
the ratio of the number of measured particles and the number of
particles incident upon the lead was found to depend only on the
distance to the shower maximum but not on the primary energy
nor on the nucleon number of the primary particle.
Therefore no systematic uncertainties in the shower size
reconstruction originate from the lead 
coverage. 
The contribution  
of muons to the number of particles
as determined with the NKG fit and the influence of the
muons on {\em age} is less than 1\% for the energy range
and detector height considered in this paper, provided that the
shower core is contained in the detector area of
180\,m$\cdot$180\,m.
Although negligible the muon component was taken into
account when simulating the pulse heights measured by the
individual scintillator huts.
Due to the conversion of photons in the lead the measured number of particles
is larger than Ne(max) for showers reaching their
maximum close to the detector.
In Figure\,\ref{eemcor} the ratio of {\em $N_e$} to
{\rm Ne(max)}
is correlated with {\em slope} measuring the
distance between detector and shower maximum.
Using {\em slope} in this correlation instead of the
distance permits the handling of all primary energies in
one correlation:
\begin{itemize}
\item The function $\xi$\sc{dis}({\em slope}) is fitted to the 
correlation in Fig.\,\ref{eemcor} and applied to
determine Ne(max) from {\em $N_e$} at detector level
and from {\em slope}.
\end{itemize}
No systematic differences between different primary particles
were noticed.
The shower size at the maximum can be reconstructed from {\em $N_e$} and
{\em slope} with accuracies ranging from 20\% (15\%) for Fe (p)
induced showers of 300\,TeV energy to 8\% (12\%) at 5\,PeV.
The accuracy for 300\,TeV iron showers is limited by the number
of fired scintillator huts, whereas
the accuracy for primary protons is always
limited by fluctuations in their shower development profiles for the
energy range and the experimental setup considered here.
In the second step the primary energy is determined from Ne(max)
and E/A.
Two sub steps are necessary here:
first the energy contained in the electromagnetic part of the EAS
has to be derived,
followed by an E/A dependent correction to
determine the primary energy.
The electromagnetic energy is
proportional to Ne(max) for a fixed shape of
the longitudinal shower development.
Because the shape changes slightly with primary energy E an iterative
procedure has to be applied to determine E finally.
Two functions are defined to  determine the primary energy from
Ne(max) and E/A.
Note that arbitrary factors may be
multiplied to
$\xi$\sc{lon}(E) and $\xi$\sc{em}(E/A) below
as long as their product is kept constant.
\begin{itemize}
\item The function $\xi$\sc{lon}(E) takes into account the change of
the longitudinal shower development with E.
It was determined from the EAS simulations:
\begin{equation}
{\rm \xi \scm{lon}(E) \, =\,
\frac{Ne(max)}{E\cdot \xi \scm{em}(E/A)}\,=\,
532\,TeV^{-1} \cdot \left[ 1. + \left( \frac{E}{6.62\,TeV} \right)
^{-0.602}\right].}
\label{eqmax}
\end{equation}
The function varies by 8\% between 300\,TeV and 5\,PeV primary energy.
\item The function $\xi$\sc{em}(E/A) is used to derive
from the electromagnetic energy and E/A the total energy E of the
primary nucleus
and will be discussed a little more detailed below.
\end{itemize}
To determine $\xi$\sc{lon} and $\xi$\sc{em} in an iterative manner
first $\xi$\sc{lon}$\equiv 1$ is assumed and $\xi$\sc{em} derived 
from a fit to the generated events.
Afterwards $\xi$\sc{lon} is fitted to the correlation of 
Ne(max)/(E$\cdot \xi$\sc{em}) versus E, which in turn 
allows to determine a new parameterization for $\xi$\sc{em}.
After two iterations neither $\xi$\sc{lon} nor $\xi$\sc{em}
change anymore. \\ 
The fraction of the primary energy which goes into the
electromagnetic part of the shower (parameterized by $\xi$\sc{em})
rises with
increasing energy as the probability for hadrons to
perform subsequent interaction with the production of
additional neutral pions (feeding the em.\ cascade by their decay
to two photons) increases with the
hadron energy.
For a nucleus the fraction of the electromagnetic
energy should depend
only on E/A.
Following the results
of \cite{eem} a function was fitted to the
correlation of
Ne(max)/E with E/A
using the generated MC events:
\begin{equation}
{\rm \xi\scm{em}(E/A)\,=\,
\frac{Ne(max)}{E \cdot \xi\scm{lon}(E)}\,=\,
\left[ 1.- \left( \frac{E/A}{33\,GeV} \right)
^{-0.181}\right]\,TeV^{-1}\ .}
\label{eqgab}
\end{equation}
The ratio of this correction
for proton to iron showers at
300\,TeV amounts to 1.34 decreasing to 1.16 at 5\,PeV
(see Figure\,\ref{lgab}).
These ratios are larger than derived by
extrapolating the results in \cite{eem}
to the mean atomic number of air because
the fraction of the total energy deposited in
the electromagnetic cascade
is different in air showers compared to
showers developing in solid state calorimeters:
in air the interaction length for charged pions
is comparable to their decay length so that
the competition between pion decay and
secondary interaction with subsequent production of
neutral pions lowers the
fraction of energy deposited in the electromagnetic cascade.\\
$\xi$\sc{em}(E/A) is determined as $\xi$\sc{em}(maximum depth), which 
is calculated as $\xi$\sc{em}({\em slope,$\theta$},E), where 
$\theta$ denotes the zenith angle.
The primary energy enters, because of the small energy 
dependence of the shape of the longitudinal shower 
development behind the maximum.
Now the primary energy is calculable by
\begin{equation}
{\rm E\,=\,\frac{Ne(max)}{\xi\scm{lon}(E) \cdot \xi\scm{em}(E/A)}
\,=\,\frac{{\em N_e}}{\xi\scm{dis}({\em slope}) \cdot 
\xi\scm{lon}(E) \cdot \xi\scm{em}({\em slope,\theta},E)}\ .}
\label{eqenemax}
\end{equation}
Due to the energy dependencies of ${\rm \xi\scm{lon} }$
and $\xi$\sc{em},
both due to a slightly changing shape of
the longitudinal shower development,
energy and distance to the shower maximum cannot strictly be determined
separately but have to be calculated iteratively.
Usually two iterations turn out to be sufficient.\\
The application of the whole procedure to simulated events
results in systematic uncertainties in the order of 5\%
for the reconstructed primary energy compared to the generated
energy (Figure\,\ref{etotaccu}).
Several contributions to the energy resolution
for iron and proton showers
are listed in Table\,\ref{tabene}.
For 300\,TeV 
\begin{table}[bht]
\caption{ The energy resolutions achieved
by using
different parameters, which were taken either directly
from the MC generator or reconstructed
from the detector information.
In the first two rows the identity of the primary
particle is used from the simulations.
``Ne(max)'', ``Ne at detector'' and ``distance''
(the distance between detector and shower maximum)
are quantities taken from the MC generator,
{\em $N_e$} and {\em slope} experimental observables. E/A denotes
the energy per nucleon, which in the last line of the table is
reconstructed from {\em slope} and the zenith angle.}
\begin{center}
\begin{tabular}{|l|l||r|r||r|r|} \hline
\multicolumn{2}{|c||}{Input Parameters} & \multicolumn{4}{|c|}{Energy Resolution (RMS)}
\\ \hline
MC gen. & Reconst. & Fe 300\,TeV &
Fe 5\,PeV & Prot.\ 300\,TeV
& Prot.\ 5\,PeV \\ \hline \hline
Ne(max) & & $(6 \pm 1)$\% & $(4 \pm 1)$\% & $(15\pm 1)$\% &
$(9 \pm 1)$\% \\ \hline
Ne at & & & & & \\
detector, & & & & & \\
distance & &$ (12 \pm 1)$\%  & $(6\pm 2)$\% & $(17\pm 1)$\% &
$(7\pm 1)$\% \\ \hline
distance & {\em $N_e$} & $(20\pm 2)$\% & $(7 \pm 2)$\% & $(18\pm 1)$\% &
$(11\pm 2)$\% \\ \hline
E/A & {\em $N_e$, slope} & $(22\pm 2)$\% & $(7\pm 2)$\% & $(15\pm 1)$\% &
$(11\pm 2)$\% \\ \hline \hline
& {\em $N_e$, slope} & & & & \\ 
& E/A & $(31\pm 3)$\% & $(12 \pm 4)$\% & $(25\pm 2)$\% &
$(11\pm 2)$\% \\ \hline
\end{tabular}
\label{tabene}
\end{center}
\end{table}
\newpage
iron showers most of the uncertainties
stem from {\em $N_e$}, determined with the NKG
fit to
a relatively small number of fired scintillator huts,
and from a modest resolution for E/A.
For energies of 1\,PeV and larger the resolution amounts
to roughly 10\%.
The much improved energy resolution is
achieved due to better {\em $N_e$} and E/A determinations and
a smaller E/A dependent correction.
The energy resolution for proton showers
improves from 25\% at 300\,TeV to about 10\% at 5\,PeV.
Above 1\,PeV even a direct measurement of Ne(max) and an unambiguous
identification of the proton would not
improve the energy resolution very much compared to the reconstruction
using only experimental observables.
\subsection{Energy Reconstruction from
Cherenkov Light alone}
\label{ecl}
Similar to the method described in the previous section the energy can be
reconstructed by replacing
{\em $N_e$} by {\em L(20-100)}.
A difference arises,
because {\em Ne} is a good measure of all charged particles
produced in e.m.\ interactions and reaching the detector level,
while {\em L(20-100)} comprise only about 20\% to 60\%
of all Cherenkov photons at detector level (Fig.\ref{complight}).
The fraction of {\em L(20-100)} compared to all 
Cherenkov photons reaching the detector level depends on the
distance to the shower maximum like {\em $N_e$},
but in addition it depends on the lateral
spread of the air shower.
This introduces an additional E/A dependency
(besides the E/A dependent fraction of the primary
energy deposited in the electromagnetic cascades),
because the lateral spread of an EAS decreases with
decreasing ratio of transverse to longitudinal momentum
in the hadronic interactions.
This effect was already discussed in the
text related to Fig.\,\ref{complight},
while comparing proton and iron showers of the same
primary energy.
The following effects were taken into account
for the energy reconstruction with Cherenkov light only:
\begin{itemize}
\item The fraction of Cherenkov light measured in
the range of 20 to 100\,m core distance {\em L(20-100)}
compared to all Cherenkov photons reaching the detector level
depends on the distance to the shower maximum
and in addition on the primary energy
due to small differences in the longitudinal
shower shapes.
Both dependencies were parameterized
with the functions
$\zeta$\sc{dis}({\em slope}) and $\zeta$\sc{lon}(E)
(corresponding to $\xi$\sc{dis} and $\xi$\sc{lon} of the previous section).
\item For a given distance and primary energy the number of all
Cherenkov photons contained in  {\em L(20-100)}
depends on E/A.
This was parameterized with the function $\zeta$\sc{em}(E/A).
\item The threshold for electrons to produce
Cherenkov light varies from 55\,MeV at a height of 150\,g/cm$^2$
to 24\,MeV at the detector level of 793\,g/cm$^2$.
Therefore the amount of Cherenkov light generated in the
atmosphere depends on the height of the shower maximum.
This is taken into account with $\zeta$\sc{den}(height).
\end{itemize}
The correction depending on energy per nucleon is
given explicitly below.
\begin{equation}
{\rm \zeta\scm{em}(E/A)}\,=\,
{\rm
\frac{{\em L(20-100)}}
{E \cdot \zeta\scm{dis} \cdot \zeta\scm{lon} \cdot \zeta\scm{den}}}
\, =\,
{\rm
\left[ 1.- \left( \frac{E/A}{178\,GeV} \right)
^{-0.180}\right]\,\frac{10^7}{TeV}}
\label{eqlgab}
\end{equation}
\begin{wrapfigure}[18]{r}{6.5cm}
\epsfig{file=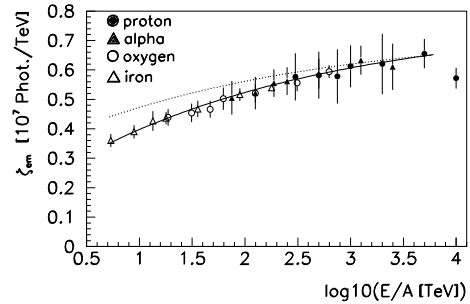,width=7cm,height=4.7cm,angle=0}
\caption{The energy per nucleon dependent
function ${\rm \zeta _{em}}$ (equation\,\ref{eqlgab}).
The error bars show the spread (RMS) for each energy
and primary.
The line shows a fit to the correlation.
The dotted line corresponds to equation\,\ref{eqgab}
normalized to eq.\,\ref{eqlgab} at 5\,PeV.}
\label{lgab}
\end{wrapfigure}

This correction varies more strongly with E/A than equation\,\ref{eqgab}
because of the discussed additional E/A dependency.
Equation\,\ref{eqgab} and \ref{eqlgab} are compared in Figure\,\ref{lgab}.
The remaining systematic uncertainties of the energy reconstruction
with Cherenkov light only
for different primary particles, energies and zenith angles
are less than 10\%,
a little worse compared to the energy reconstruction with
{\em $N_e$}.
The reason is the larger E/A dependency.
The energy resolution ranges from 45\% (35\%) at 300\,TeV
to 8\% (11\%) at 5\,PeV for iron (proton) induced showers.
Details are listed in Table\,\ref{tabecl}.
By comparing the first two rows of the table one can deduce that
the described method contains all relevant corrections
which have to be applied for an energy reconstruction.
Even optimizing
for a specific particle and energy
but restricting to Cherenkov light measurements
between 20 and 100\,m gives no improvement compared
to the algorithm of this section.
\begin{table}[bth]
\caption{The rms values of the ratio of reconstructed energy
(only from Cherenkov light)
to generated energy.
For all results the observable
{\em L(20-100)} was used.
The first row shows an ideal case,
where the energy reconstruction methods
were optimized for a special primary particle and
MC energy.
The following rows list the numbers obtained with the method described
in the text, where either MC generated or reconstructed quantities
are used.
"Distance" stands for the distance between detector and 
shower maximum.}
\begin{center}
\begin{tabular}{|l|l||r|r||r|r|} \hline
\multicolumn{2}{|c||}{Input Parameters} & \multicolumn{4}{|c|}{Energy Resolution (RMS)}
\\ \hline
MC gen. & Reconst. & Fe 300\,TeV &
Fe 5\,PeV & Prot.\ 300\,TeV
& Prot.\ 5\,PeV \\ \hline \hline
log(A), & & & & & \\
distance & {\em L(20-100)} & $(9\pm 1)$\% & $(6\pm 2)$\%
& $(15\pm 1)$\% & $(9\pm 1)$\% \\ \hline \hline
E/A, & & & & & \\
distance & {\em L(20-100)} & $(9\pm 1)$\% & $(6\pm 2)$\%
& $(15\pm 1)$\% & $(9\pm 1)$\% \\
E/A & {\em L(20-100),} & & & & \\ 
& {\em slope} & $(15\pm 2)$\% & $(6\pm 2)$\%
& $(17\pm 1)$\% & $(9\pm 1)$\% \\ \hline \hline
 & {\em L(20-100)}, & & & & \\
 & {\em slope}, E/A & $(45\pm 4)$\% & $(8\pm 3)$\%
& $(35\pm 2)$\% & $(11\pm 2)$\% \\ \hline
\end{tabular}
\label{tabecl}
\end{center}
\end{table}
Although {\em L(20-100)} can be measured (assuming a perfect 
detector) with a precision of 1\%, which is much better than a 
determination of {\em Ne} with the scintillator array, the final 
accuracy for the primary energy is even a little worse, if only
Cherenkov light is used instead of {\em Ne} and {\em slope}.
The final limitation is due to the energy contained in the non
electromagnetic shower components,
which are not measured here and have to be corrected 
for via a rather uncertain E/A estimation.
The energy determination can only be improved, if the 
hadronic and muonic EAS components are registered in addition to the 
discussed setup.  
\newpage
\section{Determination of the Chemical Composition}
\label{chemie}
\begin{wrapfigure}[14]{r}{6.5cm}     
\epsfig{file=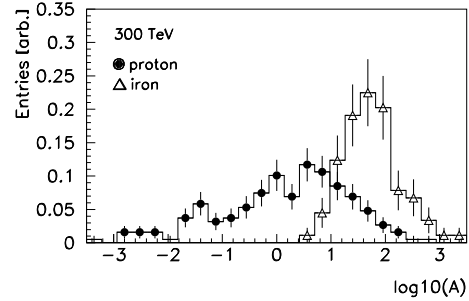,
      width=7cm, angle=0}
 \caption{The reconstructed nucleon numbers
for proton and iron primaries of 300\,TeV.}
\label{massrec}
\end{wrapfigure}
With the reconstruction of the shower properties
described in the previous sections not only the energy
is inferred independently of the mass of the
primary nucleus but also the nucleon number of the hadron hitting
the atmosphere can be determined coarsely.
In the following sections first a
nucleon number estimation derived from the longitudinal
shower development will be presented.
This procedure will be most important for the analysis
of the chemical composition.
Secondly the properties of the lateral shower extensions
will be analyzed for their sensitivity on the nucleon number of
the primary particle.
The third section combines all information concerning the
chemical composition deduced here from the four observables
{\em $N_e$, age, slope} and {\em L(20-100)}
for 300\,TeV showers as an example.
\subsection{The Chemical Composition from the
longitudinal Shower Development}
\label{complong}
Using the described procedures to determine the primary energy
and to estimate the energy per nucleon E/A
from the position of the shower maximum as described
in section\,\ref{detea}
it is straightforward to calculate the nucleon number:
\begin{equation}
{\rm log_{10}(A) \, \equiv \,
log_{10}\left(\frac{energy}{energy/nucleon}\right)
.}
\label{loga}
\end{equation}
Figure\,\ref{massrec} displays the reconstructed
log$_{10}$(A) values for proton and iron showers
of 0.3 and 5\,PeV.
The other primaries were omitted in order to keep
a clear picture.
Further results are summarized in Table\,\ref{tabmass}.
The reconstructed mean values
correspond within statistical errors to the
expectation values.
For the energy reconstruction the method with
{\em $N_e$} was used, but very similar results are obtained
with the method using only Cherenkov light to determine
the primary energy.\\
Light and heavy primaries
can be distinguished by their
different mean values and by their different spreads.
The spread of the ${\rm log_{10}(A)}$
distributions is dominated by the statistical
fluctuations of the depths of the shower maxima
with subsequent uncertainties in the E/A determination
(see Figures\,\ref{detenuc} and \ref{enucrms}).
\begin{table}[bth]
\caption{The mean and rms values of the
distributions of the reconstructed log$_{10}$(A) values.
"Rec." marks the results achieved from the reconstructed
energies and energies per nucleon as described in the text.
Differences to the fits shown in Figure\,\ref{enucrms} originate
from the summation over all zenith angles in this table.
``Gen.'' symbolizes the results obtained by using the generated
MC energy and the depth of the shower maximum directly from the
event simulation.
The numbers given in the ``Gen.'' columns therefore show the
contributions from fluctuations in the longitudinal
shower development only.
Even a perfect energy determination would hardly improve
the separation of different primary particles.}
\begin{center}
\begin{tabular}{|c||cc|c||c|c|} \cline{3-6}
\multicolumn{2}{c}{ \ }  &
\multicolumn{4}{||c|}{\rule{0cm}{0.5cm}\raisebox{0.1cm}{RMS of log$_{10}$(A)}}
 \\ \hline
Primary & \rule{0cm}{0.5cm}\raisebox{0.1cm}{$\langle$log$_{10}$(A)$\rangle$}  
& \multicolumn{1}{||c|}{300\,TeV Rec.} & 300\,TeV Gen.&
5\,PeV Rec.& 5\,PeV Gen.\\ \hline \hline
Proton & 0.00 & \multicolumn{1}{||c|}{$1.19 \pm 0.09 $ } & $1.00 \pm 0.07 $
& $0.82 \pm 0.13 $ & $0.77 \pm 0.12 $ \\
Helium & 0.60 & \multicolumn{1}{||c|}{$0.83 \pm 0.08 $ } & $0.70 \pm 0.07 $
& $0.42 \pm 0.12 $ & $0.30 \pm 0.09 $ \\
Oxygen & 1.20 & \multicolumn{1}{||c|}{$0.63 \pm 0.08 $ } & $0.52 \pm 0.07 $
& $0.26 \pm 0.08 $ & $0.22 \pm 0.06 $ \\
Iron   & 1.75 & \multicolumn{1}{||c|}{$0.53 \pm 0.06 $ } & $0.35 \pm 0.04 $
& $0.38 \pm 0.11 $ & $0.29 \pm 0.09 $ \\ \hline
\end{tabular}
\label{tabmass}
\end{center}
\end{table}
\subsection{Composition Analysis from the
lateral Shower Development}
\label{complat}
\begin{figure}[tb]
     \begin{center}
     \epsfxsize=13cm
     \epsfysize=6.16cm
     \mbox {\epsffile{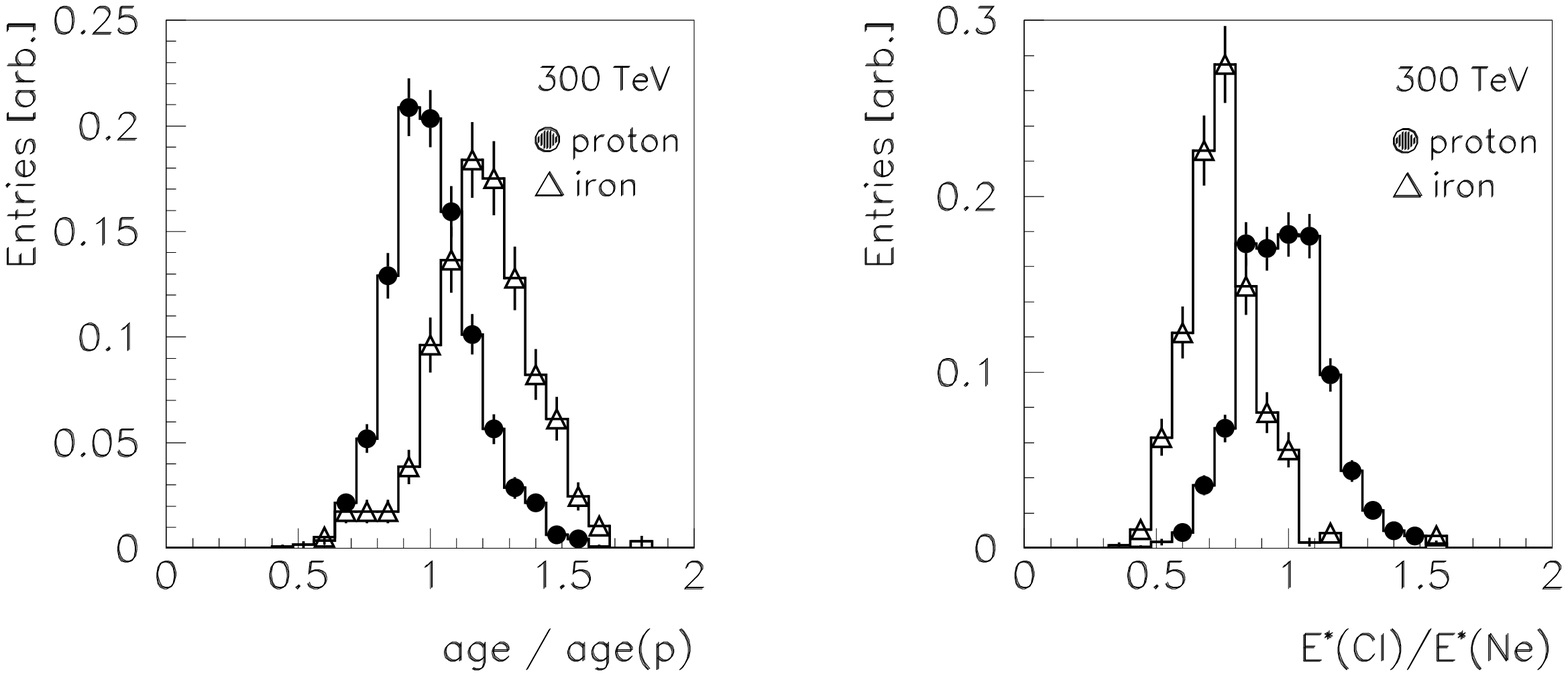}}
     \caption{The {\em age} values
divided by the expectation value for protons (left), and
the ratio of the energies reconstructed only with
Cherenkov light and by using {\em $N_e$} and {\em slope}, where
all corrections depending on E/A were omitted (right).
Proton and iron showers of 300\,TeV are displayed.
The simulated showers have been used several times
with different core positions
to take into account experimental uncertainties
related to the NKG fit.}
     \label{compae}
     \end{center}
   \end{figure}

For energies below the knee region
differences concerning the lateral EAS extensions 
(section\,\ref{showerlat}
can be used to distinguish between different primaries,
if the
energy of the primary particle and the distance to the shower maximum
are known.
The sensitive parameters are {\em age} and the fraction of 
{\em L(20-100)} compared to all Cherenkov photons at detector 
level.\\
To exploit this the expectation value of
{\em age } for proton induced showers was parameterized as
${\rm { age}(p)\, =\, 1.42 - 0.10 \cdot log_{10}(E/TeV)
+ 18.0 \cdot {\em slope}.}$
For each reconstructed shower the actual {\em age} is then
compared to the expectation for primary protons.
Figure\,\ref{compae} (left) visualizes
the distinction between 300\,TeV proton and iron induced showers.
As expected Fe showers appear to be broader than p showers.\\
When comparing the number of Cherenkov photons within and beyond
100\,m core distance, one encounters a technical difficulty,
because it is very problematic to measure the low density Cherenkov light
up to a few hundred
meters distance from the shower core with great precision.
However using the energy reconstruction methods
developed in this paper an indirect measurement of
the fraction of {\em L(20-100)} compared to all Cherenkov photons
reaching the detector level is possible:
if the energy is reconstructed only with Cherenkov light
an E/A dependent correction (eq.\,\ref{eqlgab}) has to
be applied to take into account the changing fraction of
Cherenkov light measurable within 100\,m
core distance and the fraction of
the primary energy deposited in the electromagnetic cascade.
Only the latter point has to be corrected for if the energy
reconstruction is done with the help of {\em $N_e$} (eq.\,\ref{eqgab}).
Therefore omitting all E/A corrections in both energy
reconstruction methods (resulting in E$^*$(Cl) and E$^*$({\em $N_e$}))
and then comparing E$^*$(Cl)/E$^*$({\em $N_e$})
provides an indirect estimation of the fraction of
Cherenkov light beyond 100\,m core distance.
This is equivalent to comparing the number of Cherenkov photons
between 20 and 100\,m core distance to {\em $N_e$}
taking into account the distance to the shower maximum and density
effects for the production of Cherenkov light.
In Figure\,\ref{compae} (right) the energy ratios are plotted.
A clear separation
is visible for 300\,TeV showers.
Unfortunately at energies in the "knee" region the lateral extensions
no longer depend on the primary mass
in a measurable way.
\subsection{Chemical Composition from a combined
Analysis of the longitudinal and lateral
Shower Development }
The sample of 300\,TeV showers was used to compare the
sensitivity of the different parameters discussed in the last
two sections on the mass of the primary nucleus.
If cuts in the log${\rm_{10}}$(A), {\em age}/age(p) or
E$^*$(Cl)/E$^*$({\em $N_e$}) distributions are applied
so that 90\% of the iron showers are selected, the following fractions
of proton showers remain:
\begin{itemize}
\item cut in log$_{10}$(A): 20\%,
\hspace*{1cm} -- cut in {\em age}/age(p) or
E$^*$(Cl)/E$^*$({\em $N_e$}): 40\%.
\end{itemize}
The most sensitive
parameter clearly is derived from the longitudinal
shower development.
To combine the information from the longitudinal shower
development and the lateral
extension of the EAS the probability densities,
dubbed ``$\varrho$'' below,  for observing
a specific log$_{10}$(A), E$^*$(Cl)/E$^*$({\em $N_e$})
or {\em age/age(p)} value were
parameterized for primary p, ${\rm \alpha}$, O and Fe nuclei of
300\,TeV from the MC library.
Following equation\,\ref{eqallco}
(similar for other primaries) a combined probability
${\cal P}$ is calculated.
$${\rm With \ \rho_{Fe}\, =\,\varrho({\em log_{10}(A)},Fe) \cdot
\varrho({\em E^*(Cl)/E^*(N_e)},Fe) 
\cdot \varrho({\em age/age(p)},Fe)\ :}$$
\begin{equation}
{\rm {\cal P}(Fe)\,=\,
\frac{\rho_{Fe}}{\rho_{p}+\rho_{\alpha}+\rho_{O}+\rho_{Fe}}}
\label{eqallco}
\end{equation}
Table\,\ref{taballco} lists the fractions for nuclei of
different masses which are obtained by selecting 90\% or 50\% of all
proton or iron showers.
\begin{table}[htb]
\caption{The remaining fraction of primaries with energies of 300\,TeV
after selecting
90\% or 50\% of the proton or iron showers with cuts in ${\cal P}$(p)
or ${\cal P}$(Fe).}
\begin{center}
\begin{tabular}{|l||r|r||r|r|} \hline
Primary & \multicolumn{1}{|c|}{Sel.\ 90\%\,p}
& \multicolumn{1}{|c||}{Sel.\ 50\%\,p}
& \multicolumn{1}{|c|}{Sel.\ 90\%\,Fe}
& \multicolumn{1}{|c|}{Sel.\ 50\%\,Fe}
\\ \hline \hline
Proton & 90\% \hspace*{0.8cm} & 50\% \hspace*{0.8cm}
& 8\% \hspace*{0.8cm}& $<$1\% \hspace*{0.8cm}  \\
Helium & 80\% \hspace*{0.8cm}& 17\% \hspace*{0.8cm}
& 12\% \hspace*{0.8cm}& 1.5\% \hspace*{0.8cm}\\
Oxygen & 43\% \hspace*{0.8cm}& 3\% \hspace*{0.8cm}
& 60\% \hspace*{0.8cm}& 24\% \hspace*{0.8cm}\\
Iron   & 5\% \hspace*{0.8cm}& $<$1\% \hspace*{0.8cm}
& 90\% \hspace*{0.8cm}& 50\% \hspace*{0.8cm}\\ \hline
\end{tabular}
\label{taballco}
\vspace*{0.5cm} \\
\end{center}
\end{table}
Clearly an analysis of the chemical composition
improves, if measurements of the
longitudinal and lateral shower developments are combined
at 300\,TeV.
Light and heavy particles can be separated rather well.
With the four observables used here nuclei with
masses similar to oxygen can be separated from light and heavy
primaries only in a statistical sense
but not on an event by event basis.
It seems to be difficult to distinguish between
primary protons and ${\rm \alpha}$ particles.
\section{Systematic Uncertainties}
\label{wo}
Studies of systematic effects
related to the CORSIKA code
in its version\,4.01, the influence of atmospheric transmission
as well as 
the influence of different theoretical models on the 
method described so far are addressed very briefly.
\subsection*{Effects related to CORSIKA\,4.01}
After nearly finishing the simulations used for the studies
presented here it was noticed that the EGS stepwidth used 
in CORSIKA\.4.01 is too large for simulations including 
Cherenkov light.
This results in a too narrow lateral Cherenkov light distribution
and a small overestimation of the total amount of the
Cherenkov light \cite{heckegs}.
Unfortunately the available CPU power was not
sufficient to recalculate the whole MC library.
However new simulations with the correct stepwidth only revealed 
small differences. 
Especially the presented method itself does not change in principle,
only some parameterizations have to be adopted.
This also holds for different atmospheric models which were used 
to simulate the Cherenkov light emission.
Detailed comparisons of atmospheric models
can be found in \cite{dieter}.
If a realistic atmospheric transmission of the Cherenkov light 
and a spectral response of the detectors are included in the 
simulations (which has not been done up to now to focus on the 
shower development in the atmosphere and because the detector 
efficiencies may vary for different experiments) 
the presented method does not have to be altered
principally.
One only has to
account for the zenith angle dependent detection efficiency of the 
Cherenkov light.
\subsection{Theoretical Uncertainties}
Theoretical uncertainties concerning the high energy interactions 
of primary cosmic rays in the atmosphere on the method to determine
the primary composition and energy spectrum of CR are summarized
into two categories: first the fragmentation of the primary nucleus
and afterwards different simulation models of high energy interactions
are discussed.
\subsubsection*{Maximal versus minimal Fragmentation}
To generate the showers discussed up to now the
``maximal fragmentation'' approach to model the
remnant of the primary nucleus (consisting of the
non interacting nucleons)
was used:
after the first interaction the nucleus fragments completely into
independent nucleons.
The shower maximum is reached earlier
if ``minimal fragmentation''
(the noninteracting nucleons of the primary particle
proceed as one nucleus further down the
atmosphere) is used instead, because
the cross section of a complex nucleus is larger than that of
a single nucleon.
The mean position of the maximum rises by
${\rm 22 \pm 7 \, g/cm^2}$ for 1\,PeV iron showers.
As the cross sections remain roughly constant with energy
this shift is not expected to change drastically in the
energy range covered by this paper.
Extrapolating the results for iron to oxygen and helium
with a simple geometrical parameterization yield expected
shifts of 16 and 4\,${\rm g/cm^2}$ respectively.
In addition to the shift also the spread (RMS) of the
position of the shower maximum rises slightly
if minimal fragmentation is used.
Only small differences in the shape of the longitudinal
shower development were found
not measurable with a HEGRA like detector.
The shift of the shower maximum positions for iron induced EAS
results in overestimations of the reconstructed energy 
by 20\% and of ${\rm log_{10}(A)}$ by 0.4 
if parameterizations achieved from simulations
with ``maximal fragmentation'' are applied to ``minimal fragmentation''
events. 
The overestimation of ${\rm log_{10}(A)}$ should be measurable 
with a HEGRA type of detector. 
In reference\,\cite{ost} the authors conclude that above an
energy per nucleon of 2\,GeV
the fragmentation remains scale invariant.
Therefore a measurement of
${\rm log_{10}(A)}$ for heavy primaries
well below the ``knee'' (where the results of ground based
measurements of the chemical composition can be compared
to balloon experiments)
will be sufficient to infer the fragmentation mechanism
realized in nature.
\subsubsection*{Different Interaction Models}
It goes beyond the scope of this paper to review all different
theoretical approaches to model high energy interactions.
Instead a comparison of simulations with different
interaction models
which was performed by the CORSIKA authors 
using the models DPMJET-II, HDPM, QGSJET,
SIBYLL and VENUS \cite{fzka5828}
and kindly passed to the author \cite{heckpriv}
was analyzed regarding the sensitivity of the presented methods.
These simulations do not include the emission of Cherenkov light.
Therefore the procedure to reconstruct the primary energy and mass
could not be applied directly, but other shower characteristics 
important for the described methods had to be tested.
These were the depth of the shower maxima and elongation rates,
the shape of the longitudinal shower development, the shower 
size at the maximum and the lateral shower development.
Please contact the author for details of the comparisons. 
All the different models of high energy interactions in the atmosphere
do not show drastic differences concerning the shower characteristics
used in the method described in this paper.
This is in contrast to observables related to the hadronic part of EAS.
At least up to 1\,PeV an exploration of the chemical composition
and the energy spectrum can be performed in a nearly
model independent manner,
if only observables related to the electromagnetic shower part
are used.
Further studies are needed and currently underway to investigate
model differences in the energy region of the knee and beyond.
\section{Summary and Conclusions}
In this paper methods were presented to determine energy and mass
of primary cosmic rays from ground based observations
of the electromagnetic cascade of air showers.
From the slope of the lateral Cherenkov light density in the range
of 20 to 100\,m core distance the position of the shower maximum
can be inferred without knowledge of the nucleon number of the
primary particle.
This leads to an unbiased determination of the energy per nucleon
and, combined with the shower size at
detector level or the number of
registered Cherenkov photons,
to a measurement of the primary energy.
Thus a measurement of the energy spectrum
and a coarse determination of the chemical composition
are possible
without any a priori hypotheses.
With the observables considered in the present paper
the energy resolution for primary nuclei is limited to
approximately 30\% at 300\,TeV improving to 10\% at 5\,PeV
due to natural fluctuations in the shower development.
Further improvements of these results are only possible
if accurate measurements of the non electromagnetic components
of EAS are added.\\
At energies below 1\,PeV, where results from EAS measurements
can be compared to direct data from balloon flights,
the determination of the CR mass composition
from the analysis of Cherenkov light and particles at detector level of EAS
can be substantially improved by combining
the results related to the longitudinal shower
development with parameters derived from the lateral extension.
This allows for detailed tests of the described method to determine
the chemical composition and the energy spectrum of cosmic rays. \\
The main characteristic of the methods presented here is,
that it are mainly the longitudinal shower
development behind the shower maximum,
the number of particles at the maximum
and the penetration depth of the shower until it reaches the maximum
(both mean value and fluctuations),
which determine the results.
While the first two items do not vary much for
different models describing the development of air showers
the last item is more model dependent.
In order to achieve results being as model independent as possible
it is very desirable to combine the method described in this paper
with complementary measurements.
Analyses of the early stage of the shower development, of the
hadronic and muonic components of EAS or detailed studies of the shower core
may be considered for this purpose. \\
The reconstruction of the air shower parameters as discussed in this
paper can be applied also to separate photon and nucleon induced
EAS and to reconstruct 
the primary energy of photons with a
much better precision than for primary nuclei.
This will be described
in a forthcoming paper.
\ack{The
author would like to thank the HEGRA members for
their collaboration.
Especially I am very grateful to V.\,Haustein,
who performed most of the MC simulations and determined the
chemical composition of CR with a different technique,
to G.\,Heinzelmann for many detailed, constructive
proposals and improvements as well as for his general support,
and to R.\,Plaga, who pioneered the analysis of charged CR
in the HEGRA collaboration, for motivations and detailed discussions.
Special thanks to the authors of CORSIKA for
supplying us with the simulation program and their support.
I thank Gerald Lopez for his careful reading of the text and
for providing valuable suggestions.
This work was supported by the BMBF (Germany) under contract number
05\,2HH\,264.}


\begin {thebibliography}{900}

\bibitem{shocks}
Drury,\,L.O'C., Rep.Prog.Phys.{\bf 46}, 973 (1983) \\
Blanford,\,R.D., Eichler,\,D., Phys.Rep.{\bf 154}, 1 (1987)

\bibitem{theo}
Hillas,\,A.M., Ann.Rev.Astron.Astrophys.{\bf 22}, 425 (1984) \\
V\"olk,\,H.J., Proc.\,of the ${\rm 20^{th}}$ ICRC Moscow {\bf 7}, 157 (1987) \\
Cesarsky,\,C.J., Nucl.Phys.B\,(Proc.Suppl){\bf 28B}, 51 (1992) \\
Bierman,\,P.L., ${\rm 23^{th}}$ ICRC Calagary,
Inv.\,,\,Rapp.\,and\,Highl.\,Pap., 45 (1993) \\
Biermann,\,P.L., Astron.Astrophys.{\bf 271}, 649 (1993) \\
Stanev,\,T., Biermann,\,P.L., Gaisser,\,T.K.,
Astron.Astrophys.{\bf 274}, 902 (1993) \\
Drury,\,L.O'C., Aharonian,\,F., V\"olk\,H.J.,
Astron.Astrophys.{\bf 287}, 959 (1994)

\bibitem{fzka5828}
Knapp,\,J., Heck,\,D., Schatz,\,G., FZKA Report {\bf 5828} (1997), \\
and references therein.

\bibitem{thprobs}
Gaisser,\,T.K., 
Nucl.\,Phys.\,{\bf B} (Proc.\,Suppl.) {\bf 52B}, 
10 (1997)

\bibitem{exrev}
Wdowczyk,\,W., J.Phys.\,{\bf G20}, 1001 (1994)\\
Kalmykov,\,N.N., Khristiansen,\,G.B., J.Phys.\,{\bf G21}, 1279 (1995)
Watson.\,A.A., Rapp.\,Talk ${\rm 25^{th}}$ ICRC Durban (1997)

\bibitem{kascade}
Klages,\,H.O., et al.\ (KASCADE-Collab.),\\
Nucl.\,Phys.\,{\bf B} (Proc.\,Suppl.) {\bf 52B}, 
92 (1997) \\
Kampert,\,K.-H., et al.\ (KASCADE-Collab.), ASTRO-PH/9703182 (1997)

\bibitem{hegra}
Karle,\,A., et al.\ (HEGRA-Collab.), Astropart.Phys.\,{\bf 3}, 321 (1995) \\
Krawczynski,\,A., et al.\ (HEGRA-Collab.), NIM\,{\bf A383}, 431 (1996) \\
Lindner,\,A., et al.\ (HEGRA-Collab.),\\
Proc.\,of the ${\rm 25^{th}}$ ICRC Durban {\bf 5}, 113 (1997)

\bibitem{perpig}
Aharonian,\,F., Heinzelmann,\,G., Proc.\ of the {15${\rm ^{th}}$ ECRS},
Perpignan (1996), in press (similar in ASTRO-PH/9702059)

\bibitem{romax}
Lindner,\,A., et al.\ (HEGRA-Collab.), \\
Talk at the 24${\rm ^{th}}$ ICRC, Rome (1995), unpublished \\
Cortina, J., et al.\ (HEGRA-Collab.),\\
Proc.\,of the ${\rm 25^{th}}$ ICRC Durban {\bf 4}, 69(1997)

\bibitem{romra}
Plaga,\,R., et al.\ (HEGRA-Collab.),
Proc.\,of the ${\rm 24^{th}}$ ICRC Rome {\bf 2}, 693 (1995)

\bibitem{oldcheren}
Protheroe,\,R.J., Turver,\,K.E., Nuovo.Cim.\,{\bf 51A}, 277 (1979) \\
Chantler,\,M.P., et al.\ J.Phys.\,{\bf G8}, L51 (1982) \\
Turver,\,K.E., Nucl.Phys.B\,(Proc.Suppl.)\,{\bf 28B}, 16 (1992)\\
Kalmykov,\,N.N., Khristiansen,\,G.B., Prosin,\,V.V.,\\ 
Phys.At.Nucl.\,{\bf 58}, 1563 (1995) 

\bibitem{nkg}
Kamata,\,K.,  Nishimura,\,J., Prog.Theor.Phys.,\,Suppl.\,{\bf 6}, 93 (1958)\\
Greisen,\,K. Ann.Rev.Nucl.Sci.{\bf 10},63 (1960)

\bibitem{cors}
Capdevielle,\,J.N. et al., KFK Report {\bf 4998} (1992) \\
Knapp,\,J., Heck,\,D., KFK Report {\bf 5196B} (1993)

\bibitem{volker}
Haustein,\,V., doctoral thesis at Univ.\,of\,Hamburg (1996)

\bibitem{hillas}
Patterson,\,J.R., Hillas,\,A.M., J.Phys.\,{\bf G9},1433 (1983)

\bibitem{eem}
Gabriel,\,T.A., et al., Nucl.Instrum.Meth.\,{\bf A338},336 (1994)

\bibitem{heckegs}
Heck,\,D., private communication

\bibitem{dieter}
Horns,\,D. diploma thesis, Univ.\ of Hamburg (1997)

\bibitem{ost}
Kalmykov,\,N.N. and Ostapchenko,\,S.S., Phys.At.Nucl. {\bf 56},
346 (1993) 

\bibitem{heckpriv}
Heck,\,D., private communication

%
\end{thebibliography}
\end{document}